%% file: main.tex
\documentclass[sigconf, nonacm]{acmart}

\newcommand\vldbdoi{XX.XX/XXX.XX}
\newcommand\vldbpages{XXX-XXX}
\newcommand\vldbvolume{14}
\newcommand\vldbissue{1}
\newcommand\vldbyear{2021}
\newcommand\vldbauthors{\authors}
\newcommand\vldbtitle{\shorttitle} 
\newcommand\vldbavailabilityurl{http://vldb.org/pvldb/format_vol14.html}
\newcommand\vldbpagestyle{plain} 

\usepackage{pgf}
\usepackage{algorithm}
\usepackage{algorithmicx}
\usepackage{algpseudocode}
\usepackage{subfigure}

\algnewcommand\algorithmicforeach{\textbf{for each}}
\algdef{S}[FOR]{ForEach}[1]{\algorithmicforeach\ #1\ \algorithmicdo}

\algblockdefx{ForAllP}{EndFaP}[1]%
  {\textbf{for all}#1 \textbf{do in parallel}}%
  {\textbf{end for}}
\algtext*{EndFaP}%

\algblockdefx{ForAllNP}{EndFaP}[1]%
  {\textbf{for all}#1 \textbf{do serially}}%
  {\textbf{end for}}
\algtext*{EndFaP}%

\algnewcommand\algorithmicswitch{\textbf{switch}}
\algnewcommand\algorithmiccase{\textbf{case}}
\algnewcommand\algorithmicassert{\texttt{assert}}
\algnewcommand\Assert[1]{\State \algorithmicassert(#1)}%
\algdef{SE}[SWITCH]{Switch}{EndSwitch}[1]{\algorithmicswitch\ #1\ \algorithmicdo}{\algorithmicend\ \algorithmicswitch}%
\algdef{SE}[CASE]{Case}{EndCase}[1]{\algorithmiccase\ #1}{\algorithmicend\ \algorithmiccase}%
\algtext*{EndSwitch}%
\algtext*{EndCase}%

\newcommand{\rmi}{RMI}
\newcommand{\rs}{RadixSpline}

\newcommand{\fsm}{FSM}
\newcommand{\amac}{AMAC}

\newcommand{\simd}{SIMD}
\newcommand{\avx}{AVX512}

\usepackage{xcolor}

\begin{document}
\title{When Are Learned Models Better Than Hash Functions? (Extended Abstracts)}

\newcommand{\kapil}[1]{\stepcounter{todocounter}
  {\color{magenta!90} Kapil: \thetodocounter: #1}} 

\author{Ibrahim Sabek}
\authornote{Both authors have equal contributions and their names are sorted alphabetically.}
 \affiliation{
   \institution{MIT CSAIL}
 }
 \email{sabek@mit.edu}

 \author{Kapil Vaidya}
 \authornotemark[1]
 \affiliation{
   \institution{MIT CSAIL}
 }
 \email{kapilv@mit.edu}

 \author{Dominik Horn}
 \affiliation{
   \institution{TUM}
 }
 \email{dominik.horn@tum.de}

 \author{Andreas Kipf}
 \affiliation{
   \institution{MIT CSAIL}
 }
 \email{kipf@mit.edu}

 \author{Tim Kraska}
 \affiliation{
   \institution{MIT CSAIL}
 }
 \email{kraska@mit.edu}

\begin{abstract}
In this work, we aim to study when learned models are better hash functions, particular for hash-maps.
We use lightweight piece-wise linear models to replace the hash functions as they have small inference times and are sufficiently general to capture complex distributions.
We analyze the learned models in terms of: the model inference time and the number of collisions. 
Surprisingly, we found that learned models are not much slower to compute than hash functions if optimized correctly. 
However, it turns out that learned models can only reduce the number of collisions (i.e., the number of times different keys have the same hash value) if the model is able to over-fit to the data; otherwise, it can not be better than an ordinary hash function. 
Hence, how much better a learned model is 
in avoiding collisions highly depends on the data and the ability of the model to over-fit. 
To evaluate the effectiveness of learned models, we used them as hash functions in the bucket chaining and Cuckoo hash tables.
For bucket chaining hash table, we found that learned models can achieve 30\% smaller sizes and 10\% lower probe latency. 
For Cuckoo hash tables, in some datasets, learned models can increase the ratio of keys stored in their primary locations by around 10\%.
In summary, we found that learned models can indeed outperform hash functions  but only for certain data distributions and with a limited margin. 
\end{abstract}

\maketitle

\pagestyle{\vldbpagestyle}
\begingroup\small\noindent\raggedright\textbf{PVLDB Reference Format:}\\
\vldbauthors. \vldbtitle. PVLDB, \vldbvolume(\vldbissue): \vldbpages, \vldbyear.\\
\href{https://doi.org/\vldbdoi}{doi:\vldbdoi}
\endgroup
\begingroup
\renewcommand\thefootnote{}\footnote{\noindent
This work is licensed under the Creative Commons BY-NC-ND 4.0 International License. Visit \url{https://creativecommons.org/licenses/by-nc-nd/4.0/} to view a copy of this license. For any use beyond those covered by this license, obtain permission by emailing \href{mailto:info@vldb.org}{info@vldb.org}. Copyright is held by the owner/author(s). Publication rights licensed to the VLDB Endowment. \\
\raggedright Proceedings of the VLDB Endowment, Vol. \vldbvolume, No. \vldbissue\ %
ISSN 2150-8097. \\
\href{https://doi.org/\vldbdoi}{doi:\vldbdoi} \\
}\addtocounter{footnote}{-1}\endgroup


\input{introduction}
\input{background}
\input{analysis}
\input{evaluation}
\newpage
\balance


\appendix
\input{appendix}

\bibliographystyle{plain}
\bibliography{main}

\balance

\end{document}

%% file: introduction.tex
\section{Introduction}

Hashing is a fundamental operation in computer science and commonly used in databases~\cite{RAD15}. They are mainly used to accelerate point queries, perform joins and grouping, etc. (e.g.,~\cite{BTA+13, JMH+18}).
 In hash tables, a key is mapped to a location in constant time (i.e., $O(1)$).
Compared to the traditional tree-structured main-memory indexes, hash tables have been proven to be much faster for point queries.
Meanwhile, a lot of data structures and algorithms are recently being enhanced by learned models (e.g.,~\cite{KBC+18, KVC+20}). These learned structures can outperform their traditional counterparts on practical workloads. Along this line of work, the authors of~\cite{KBC+18} introduced the idea of using learned models instead of hash functions, and supported that by some empirical evidence. 
In this paper, we aim to study in more detail when learned models are better hash functions, particular for applications like hash-maps. We primarily consider piece-wise linear models in our analysis as they have small inference times and are sufficiently general to capture complex distributions.

Our study investigates the performance of learned models in terms of: \textit{the number of collisions}, and \textit{the computation time}. 
A collision between two keys occurs when they have the same hash value. Some hash functions are extremely fast to compute, yet, they might suffer from a considerable collision rate in some scenarios (e.g., Multiply-shift~\cite{DHK+97}).
In general, there are known theoretical lower bounds on the number of collisions achieved by hash functions.
We observed that learned models might be able to do better than these lower bounds and outperform hash functions. In particular, it turns out that the amount of collisions for learned models is dependent on the data distribution.

Regarding computation time, we empirically found that learned models are slower to compute than most hash functions due to the cache miss overhead from randomly accessing the model's parameters. However, with the help of vectorization and prefetching-optimized inter-task parallelism (e.g., {\amac}~\cite{KFG15}), the learned models computation time can come quite close to its hashing counterpart (around 2 ns difference using models with moderate size).


To show the effect of using learned models within hashing applications, we built bucket chaining and Cuckoo hash tables using two efficient models, namely {\rmi}~\cite{KBC+18} and {\rs}~\cite{KMR+20}, instead of hash functions.
Typically, in Cuckoo hashing, a single hash function is used to extract two hash sequences. In our experiments, we computed one hash sequence using the learned model and the other using the hash function.  
We empirically evaluate the performance of these altered hash tables with various real-world and synthetic datasets.
For bucket chaining, we found that learned models can achieve 30\% smaller hash tables and 10\% lower probe latency. 
For Cuckoo hashing, in some datasets, learned models can increase the ratio of keys stored in their primary locations (primary key ratio) by around 10\% and a small lookup time benefit in the probe phase.





%% file: background.tex
\section{Background}
\label{sec:background}

\input{background_hash_functions_and_tables}
\input{background_learned_models}
\input{background_parallelism_optimizations}

%% file: background_hash_functions_and_tables.tex
\noindent\textbf{Hash Functions and Tables.} Murmur~\cite{RAD15} and XXH3~\cite{XXHash} are among the most widely-used hash functions, which have good balance between computation time and collision rates. They are implemented with arithmetic (e.g., multiply, add) and logical (e.g., shift, XOR) operations. However, XXH3 is specifically designed for streaming data. AquaHash~\cite{AquaHash} is another popular hash function that leverages Advanced Encryption Standard (AES) instructions~\cite{AES}. In general, hashing schemes for handling collisions are categorized into two main categories: chaining and open addressing. Bucket chaining~\cite{BTA+13} is a standard hash table implementation that follows the chaining scheme. It contains a set of $n$ buckets, where each bucket has a pre-allocated array of $s$ entries. On an insert, once a collision occurs, the item is inserted in the current available entry in its corresponding bucket. If the current bucket is already filled up, a new one is created, pre-allocated and chained to it. For open addressing scheme, Cuckoo hash table~\cite{PR04} has become the recent state-of-the-art.
Every item has two possible locations: its primary and its secondary bucket.
When inserting an item and its primary bucket is full, it gets placed into its secondary bucket.
If the secondary bucket is also full, a random item is kicked from the bucket and is placed into its alternative location (balanced kicking).
In contrast, biased kicking~\cite{cuckoo-index} prefers kicking items that reside in their secondary buckets.
The idea behind this is to increase the ratio of items in their primary buckets (primary ratio) and hence improve performance for positive lookups.
Typically, Cuckoo hashing is implemented with two independent hash functions.

%% file: background_learned_models.tex
\noindent\textbf{Learned Models.} Recently, the idea of using learned models to predict the location of keys in datasets has gained a great attention in the database community~\cite{KBC+18}. {\rmi}~\cite{KBC+18} was the first proposed index that uses multi-stage learned models. In {\rmi}, the root model gives an initial prediction of the CDF for a specific key. Then, this prediction is recursively refined by more accurate models in the subsequent stages. Interestingly, the authors of~\cite{KBC+18} also discussed the idea of using CDF-based learned models as order-preserving hash functions, which is the main scope of this paper. An interesting index that followed {\rmi}, namely {\rs}~\cite{KMR+20}, employs a radix table to quickly find the two spline points that approximate the CDF for a specific key. Then, linear interpolation between the retrieved spline points is used to locate the key.  In this paper, we only focus on piece-wise linear models that are built using a set of line segments, where each segment is represented by a slope and an intercept. Both {\rmi} and {\rs} can be considered as piece-wise linear models, which are just trained/created in a different way. 

%% file: analysis.tex
\vspace{-5pt}
\section{Analysis of Learned Models}

\input{analysis_models_for_less_collisions}
\input{analysis_models_as_fast_hash_functions}

%% file: analysis_models_for_less_collisions.tex
\subsection{Can Learned Models Cause Less Collisions?}
\label{sec:analysis_collisions}

In this section, we first characterize collisions and then use this to identify/analyze factors affecting collisions for learned models and hash functions.
This analysis helps us to characterize situations where learned models outperform hash functions.

\noindent\textbf{Notation.} We consider the task of mapping $N$ keys to $N$ locations for ease of analysis.
$x_0,x_1,...$ is the sorted array of $N$ keys ($x_i<=x_{i+1}$) and $y_0,y_1,..$, where $y_i\in[0,N-1]$, is the corresponding sorted array of output values ( $y_j=f(x_i)$) where f is a learned model or hash function) such that $y_i<=y_{i+1}$. 
Note that $x_i$ does not necessarily relate to $y_i$, $x_i$'s and $y_i$'s are just sorted versions of the original input keys and output locations.
For learned models, $y_i$'s are continuous values and the precise output location is the closest integer to $y_i$'s.
The sorted output values generate a set of gaps $g_1, g_2,...$ such that $y_i=\left(\sum_{t=1}^i g_t\right)+y_0$.
These gaps form a distribution $G$ with a probability density function (PDF) $f_{G}$.
The discussion and analysis in the rest of this section support the following points:
\begin{itemize}
    \item Collisions are dependent on the gaps between consecutive sorted output values ($g_i$'s).
   
    \item For piece-wise linear models, the number of collisions is dependent on the key distribution, specifically the gaps between consecutive sorted keys ($x_{i}-x_{i-1}$). Higher variation in the distribution of gaps between keys leads to more collisions. Having more linear models improves the accuracy but may not reduce the collisions.
    \item Collisions for a good hash function are independent of the input key distribution (The distribution of $x_i$'s).
\end{itemize}


\noindent\textbf{Characterizing Collisions.}
If two keys are mapped to the same location, then there is a collision.
The key insight regarding collisions is that the collisions depend on the gaps between consecutive output values (${y_i-y_{i-1}}$).
If the gap between two consecutive output values is greater than one $(y_i-y_{i-1}\ge1)$, then they would definitely be placed in separate locations. On the other hand, if the the gap is smaller than one $(y_i-y_{i-1}\leq1)$, they may be mapped to the same location depending on the location boundary.
In addition, the smaller the gap value the more the probability of the keys falling in the same location.

The gap values are constrained by the condition that the sum of all the gaps should be less than $\left(N-1\right)$\footnote{Sum of gaps is: $\sum_{t=1}^{N-1}\left(y_t-y_{t-1}\right) =y_{N-1}-y_{0} \leq N-1$}, and thus, the mean gap value turns out to be less than or equal to one ($E[G] \leq 1$). 
Ideally, we would want all the gaps to be exactly equal to one as this leads to zero collisions and also satisfies the constraint.
Qualitatively speaking, any increase in the variance of gap distribution $G$ leads to an increase in the number of gaps below value one and thus, a subsequent increase in the number of collisions.


\noindent\textbf{Collisions for Linear Models.}
The output distribution for linear models is dependent on the input distribution.
Linear operations scale and offset the input values to obtain the output.
For sorted input values (${x_0,x_1,x_2,..}$), a simple linear model ($y=m * x+b$) will just scale the gaps of the input ($g_i=y_{i+1}-y_{i}=(x_{i+1}-x_{i})*m$).
For piece-wise linear models, the gap distribution of the output values $G$ is a scaled version of the input. The scaling will be such that the mean of $G$ is less than equal to one ($E[G]\leq 1$). 
If the input gap distribution has higher variance, this would be propagated to $G$, leading to more collisions. 

Next, we qualitatively argue why more models do not necessarily reduce the number of collisions.
Suppose the input data was generated using a gap distribution $H$ with corresponding PDF $f_H$.
Piece-wise linear models would simply scale different ranges of the input and thus, the corresponding output gap distribution would just be a scaled version of $f_H$. Increasing the number of models does not alter the gap distribution of the output values and thus, the number of collisions stays the same.
In an extreme case, when the number of models is close to the number of keys, then the collisions would be low but the space overhead would make the structure practically unusable.   

\begin{figure}
    \begin{center}
    \includegraphics[width=6cm,height=4cm]{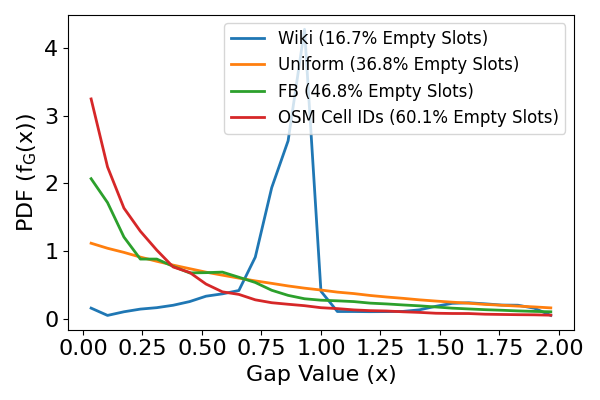}
    \end{center}
    \caption{Gap distribution and effect on collisions}
    \label{fig:gap_distribution}
\vspace{-2mm}    
\end{figure}

Here, we visualize the gap distribution of the output values for various datasets and the corresponding proportion of empty slots.
We used piece-wise linear models for various datasets and obtained the output $y_i's$ (CDF) values for them. 
In Figure~\ref{fig:gap_distribution}, we show the PDF of the gap distribution and the proportion of empty slots for three real datasets from~\cite{sosd_bench} and a synthetic uniform one.
Clearly, the gap distribution is much more predictable for \textit{wiki} than for uniform, \textit{fb} and \textit{osm}. 
\textit{wiki} has a gap distribution concentrated more towards one and so ends up having the least number of empty slots.
\textit{osm} tends to have a lot of gaps concentrated towards zero and ends up with the most empty slots.
We provide a formal analysis for the collisions and its relation with gaps distribution in Appendix~\ref{app:collisions}.

\noindent\textbf{Collisions for Hash Functions.} 
In case of a good hash function (e.g., Murmur), the output values will be  uniformly distributed in the range $[0,N-1]$ irrespective of the input distribution. 
Therefore, the gap distribution of the output values is a fixed distribution which corresponds to the uniform case in Figure~\ref{fig:gap_distribution}.

\noindent\textbf{Summary.}
If the input keys are generated from a distribution, then the CDF of the distribution maps the data uniformly randomly in the range $[0,1]$.
Hence, the CDF will behave as an order-preserving hash function in a hash table.
A learned model that approximates this underlying distribution would only be as effective as a hash function in terms of collisions.
In this case, over-fitting to the data is necessary to reduce the collisions. 
If the data is generated in an incremental time series fashion ($x_0,x_1=x_0+g_0,x_2=x_1+g_1,....$), the predictability of the gaps determines the amount of collisions. 
In certain cases, like the \textit{wiki} distribution, a learned model can lead to fewer collisions.
Auto generated IDs with some deletions are the other common case where learned models can beat hash functions. 

%% file: analysis_models_as_fast_hash_functions.tex
\subsection{Can Learned Models be Computationally as Fast as Traditional Hash Functions?}

In this section, we first present the computation bottleneck of using learned models as hash functions. Then, we discuss the opportunity of alleviating such bottleneck by using vectorization (i.e., Single Instruction Multiple Data ({\simd}) instructions) and prefetching-optimized inter-task parallelism techniques (e.g., {\amac}~\cite{KFG15}). 

\noindent\textbf{Cache Misses Overhead.} The computation of traditional hash functions is fast. It usually includes the execution of arithmetic, logical, and shifting operations (e.g., Multiply-shift~\cite{DHK+97}, and Murmur~\cite{RAD15}). 
In contrast, using learned models, like {\rmi}, as a hash function incurs higher latency. The main reason is that, although the inference computation of these models (which is basically the hashing computation in our case) is completely based on arithmetic operations (e.g., add, multiply, max), there is an additional overhead in accessing the model parameters (e.g., intercepts and slopes) that will be used during the computation. This overhead significantly increases if the model size becomes large as its parameters will not completely fit in the cache, and randomly accessing them from the memory will incur many cache misses.  

\noindent\textbf{Performance Gain via {\simd}.}
Interestingly, vectorizing the computation in learned models is \textit{more efficient} than vectorizing some hash functions \textit{as long as the model parameters are kept warm in the cache}. To backup this claim, we micro-benchmarked the throughput of hashing 128 million 64-bit integer keys using a single-threaded {\avx} {\simd} implementation for both Murmur~\cite{Murmur3} and 2-levels {\rmi} model~\cite{KBC+18}, running on a machine with Intel(R) Xeon(R) processors (Skylake architecture). 
We made sure that all models' parameters are kept in the cache by building an {\rmi} with a total of 5 linear models only (1 root model, and 4 models in the second level). Our results showed that the hashing throughputs for vectorized {\rmi} and vectorized Murmur are 1000 and 800 million keys/sec, respectively. This is expected because, \textit{with ignoring the effect of parameters' cache misses}, the inference computation (i.e., hashing) in {\rmi} heavily relies on fast comparison (e.g., min/max) and fused instructions\footnote{Intel(R) Xeon(R) Gold 6230 processor has two physical {\avx} FMA units.} (e.g., fmadd), each has a throughput of 2 instructions/cycle~\cite{IntelIntrinsics}. On the other hand, 60\% of the total instructions needed in the Murmur computation have a throughput of 1 instruction/cycle or less, such as logical shift (1 instruction/cycle) and multiplication (0.66 instruction/cycle). 

\noindent\textbf{Performance Gain via {\amac}.}
As previously mentioned, the superiority of vectorized learned models quickly diminishes when we have large models, which is a typical case in real settings. In this case, the model parameters are frequently accessed from memory and not from cache. Even if some of the requested parameters from a vectorized instruction hit in cache, the instruction cannot proceed until cache misses of the other parameters in the vector are resolved. Obviously, direct software prefetching is not a feasible solution to this issue, and will completely stall the performance, because the model parameters require immediate memory access. Therefore, we propose to hide the cache misses latency by combining the vectorized learned models with a widely-used prefetching-optimized inter-task parallelism technique, namely {\amac}~\cite{KFG15}. This helps in making the overall latency of vectorized learned models very close to traditional hash functions as shown in our evaluation (Section~\ref{sec:evaluation}).
Appendix~\ref{app:hash_via_learned_models_algo} shows our proposed batch-oriented hash function that combines the benefits of {\simd} and {\amac} with learned models.

%% file: evaluation.tex
\section{Evaluation}
\label{sec:evaluation}


For the experiments, we use three real datasets from \cite{sosd_bench}, namely \textit{wiki}, \textit{osm}, and \textit{fb}, in addition to three variations of a synthetic sequential dataset with different x\% elements removed randomly (x=\{0, 1, 10\}). 
Each real or synthetic dataset has around 200 million 64-bit keys. We de-duplicate the real datasets before using them.
We use AquaHash~\cite{AquaHash}, XXH3~\cite{XXHash} and Murmur~\cite{Murmur3} with fast modulo reduction\footnote{Modulo reduction is based on efficient integer division (https://libdivide.com).} as traditional hash functions and two efficient learned models: {\rmi}~\cite{KBC+18} and {\rs}~\cite{KMR+20}. 

\begin{figure}
     \centering
     \subfigure[Median keys per  second]{\includegraphics[width=1.6in,height=2.7cm]{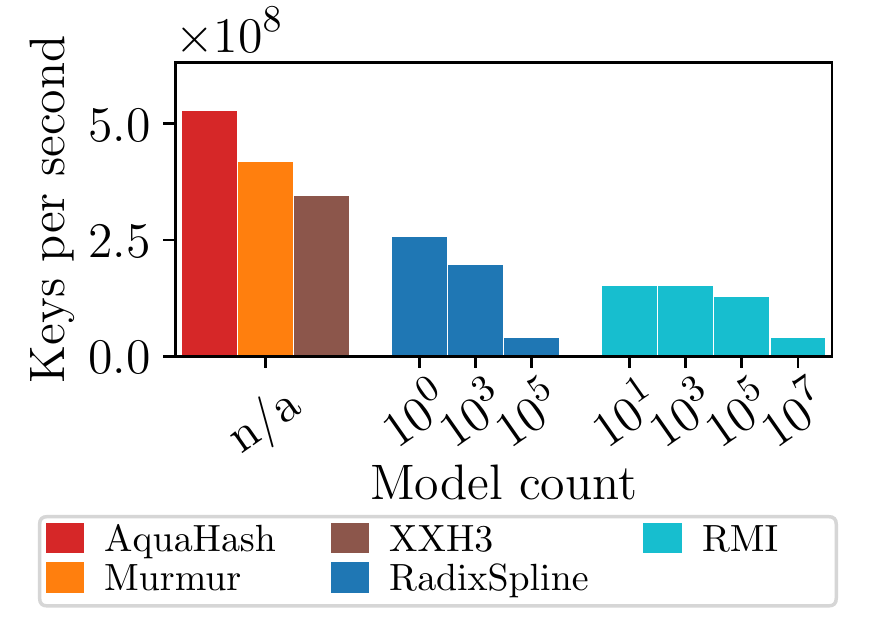}
     \label{fig:computation_time}}
     \subfigure[Empty slots (percentage)]{\includegraphics[width=1.6in]{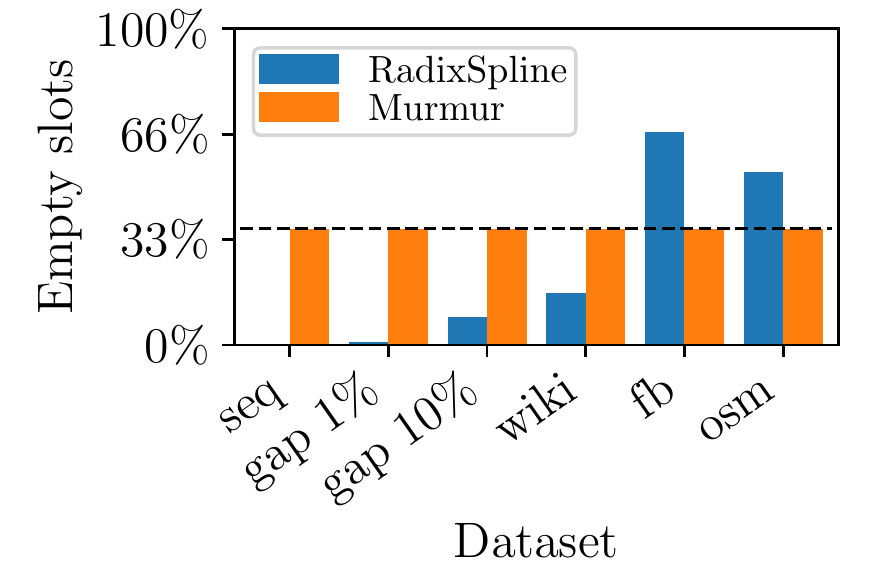}
     \label{fig:collisions}}
     \caption{Throughput and collisions comparisons.}
     \label{fig:computation_time_and_collisions}
\end{figure}

\noindent\textbf{Computation Time.} Figure~\ref{fig:computation_time} shows the median throughput for hashing the sequential dataset with 10\% removed elements using traditional hash functions and non-vectorized learned models, while varying the count of line segments used in each model. As expected, traditional functions are much better than non-vectorized learned models, even with small sizes. The throughput of learned models decreases significantly for large sizes due to the increased number of cache misses when accessing the model parameters. Table~\ref{table:rmi_compute} shows the performance of {\amac}-based vectorized versions of both {\rmi} and Murmur hashing for the same dataset. As shown, with $10^3$ models, the performance gap between non-vectorized {\rmi} and Murmur is substantial (around 10 ns difference), however, using the {\amac}-based vectorization reduces this gap to be 2 ns only. Note that, at very large models (e.g., $10^7$), {\rmi} becomes much slower than Murmur, even with vectorization.

\begin{table}
\small
\begin{tabular}{|l|l|l|l|l|}
\hline
\begin{tabular}[c]{@{}l@{}}\scriptsize{Model} \\ \scriptsize{Count} \end{tabular} & \begin{tabular}[c]{@{}l@{}}\scriptsize{Non-Vect.} \\ \scriptsize{Murmur (ns)} \end{tabular} & \begin{tabular}[c]{@{}l@{}}\scriptsize{Vect.} \\ \scriptsize{Murmur (ns)} \end{tabular} & \begin{tabular}[c]{@{}l@{}}\scriptsize{Non-Vect.} \\  \scriptsize{RMI (ns)}\end{tabular} & \begin{tabular}[c]{@{}l@{}}\scriptsize{Vect.}  \\  \scriptsize{RMI (ns)}\end{tabular} \\ \hline
10       & 2.4 & 1.9 & 10 & 4 \\ \hline
$10^3$   & 2.4 & 1.9 & 13 & 4 \\ \hline
$10^5$   & 2.4 & 1.9 & 25 & 8 \\ \hline
$10^7$   & 2.4 & 1.9 & 112 & 22 \\ \hline
\end{tabular}
\caption{Computation time of vectorized {\rmi} and Murmur.}
\label{table:rmi_compute}
\vspace{-10mm}
\end{table}

\noindent\textbf{Collisions.} Figure~\ref{fig:collisions} shows the amount of collisions for {\rs} against Murmur hashing on various datasets. {\rmi} is omitted as it has similar results to {\rs}. Here, we mapped $N$ elements into $N$ slots and report the proportion of empty slots. The dashed line is the theoretically expected value for true uniform random hash functions. As shown in the figure, for many datasets, learned models indeed have less empty slots than hash functions (i.e., less collisions). However, for \textit{fb} and \textit{osm} datasets, the models make the collisions worse. This confirms our analysis in Section~\ref{sec:analysis_collisions}.



\begin{figure}
    \begin{center}
    \includegraphics[width=3in]{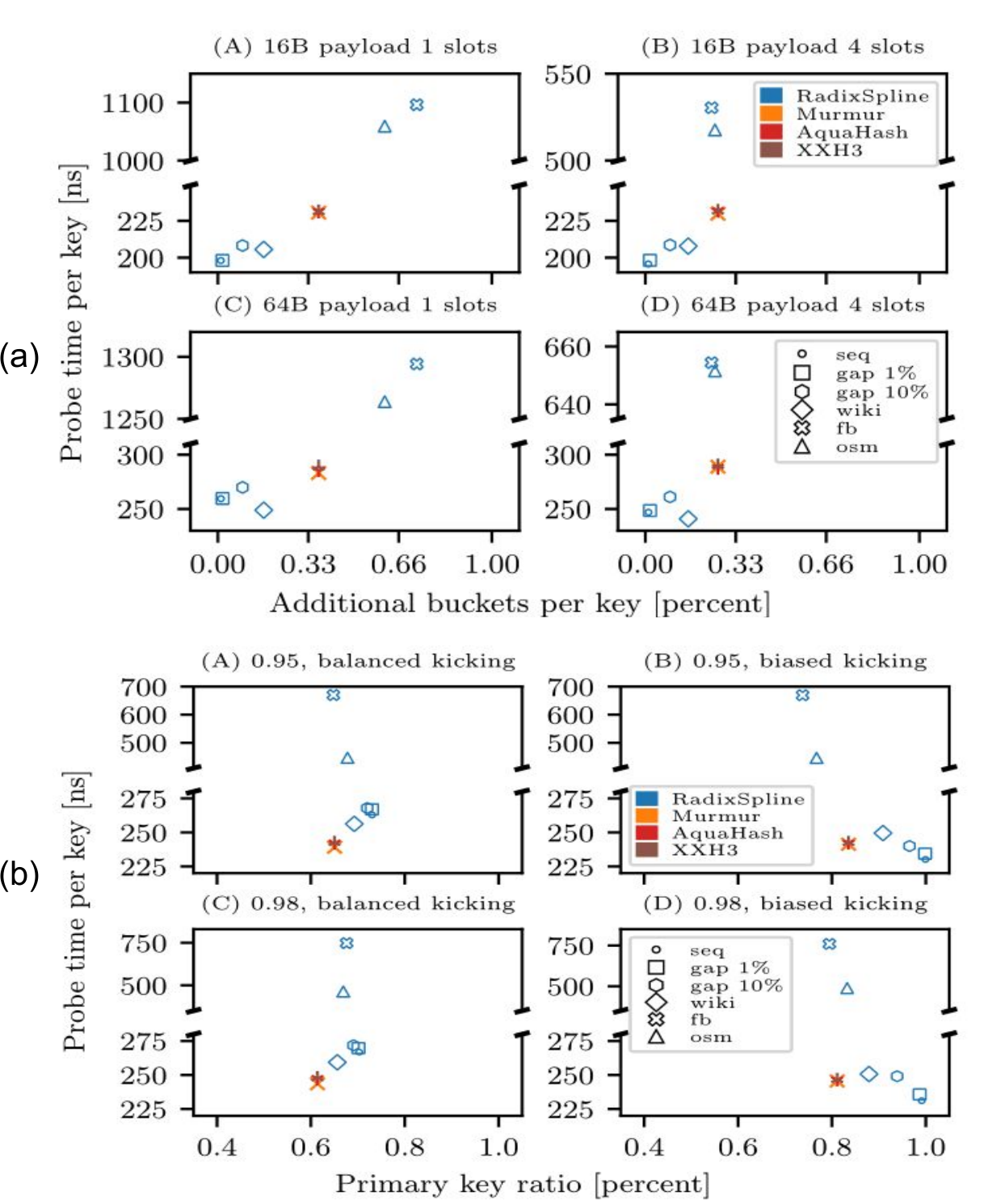}
    \end{center}
    \caption{(a)~Bucket chaining hash table probe times for varying payload sizes and slots per bucket. (b)~Primary key ratio and probe times for various kicking strategies and load factors.}
    \label{fig:primary_key}
\end{figure}


\noindent\textbf{Bucket Chaining Hash Tables.} Bucket chaining hash tables deal with collisions by creating linked lists for the keys mapped to the same location. When retrieving a key, we traverse the linked list until we find the key. With increased collisions, the space needed for the chained hash tables increases. Figure~\ref{fig:primary_key}(a) shows the effect of using different hash functions and {\rs} as a representative learned model when building bucket chaining hash tables with different payload and bucket sizes. We observe that {\rs} can lead to less probe times for all of the datasets, except \textit{fb} and \textit{osm} ones. Moreover, larger payloads lead to larger cache miss penalties, and hence with increasing payload sizes, hash functions take slightly more time than learned models.


\noindent\textbf{Cuckoo Hash Tables.} 
Having a high primary key ratio reduces the amount of unnecessary lookups, as one avoids going to the second location, and hence improves the probe time. We show the effect of replacing one of the used multiple hash functions by a learned model. We use a Cuckoo Hash with 2 hash functions, load factor of 1, bucket size of 8, and two kicking strategies; balanced and biased~\cite{cuckoo-index}. As shown in Figure~\ref{fig:primary_key}(b), using any two traditional hash functions consistently achieves primary key ratios of 62\% and 83\%, for biased and balanced kicking, respectively, which are close to theoretically optimal. However, we observe that using learned models, e.g., {\rs}, along with both kicking strategies can lead to better primary key ratio for all of the datasets, except \textit{fb} and \textit{osm} ones.
With biased kicking, learned models get a much better primary key ratio which leads to lower cache misses and thus a slightly better probe time.

\begin{figure}
     \centering
     \subfigure[16 bytes payloads]{\includegraphics[width=1.6in,height=2.7cm]{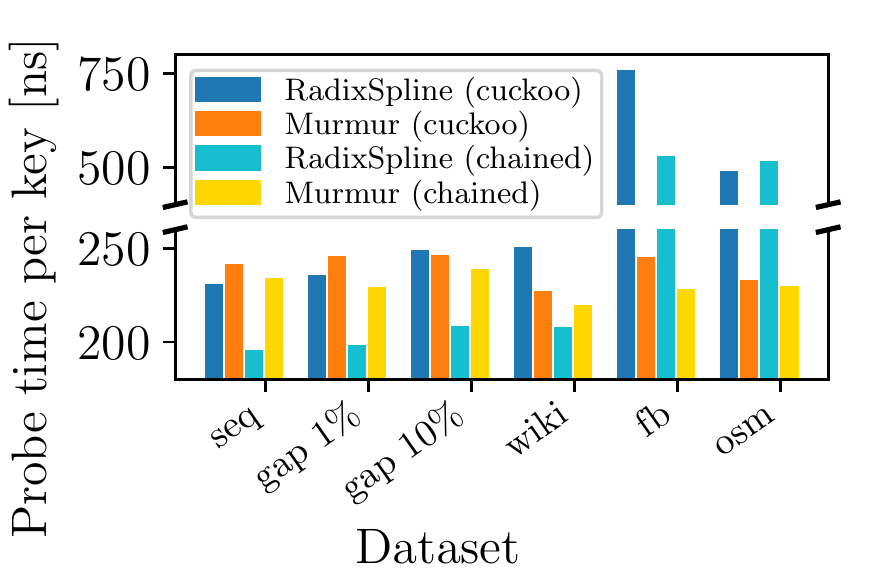}}
     \subfigure[64 bytes payload]{\includegraphics[width=1.6in,height=2.7cm]{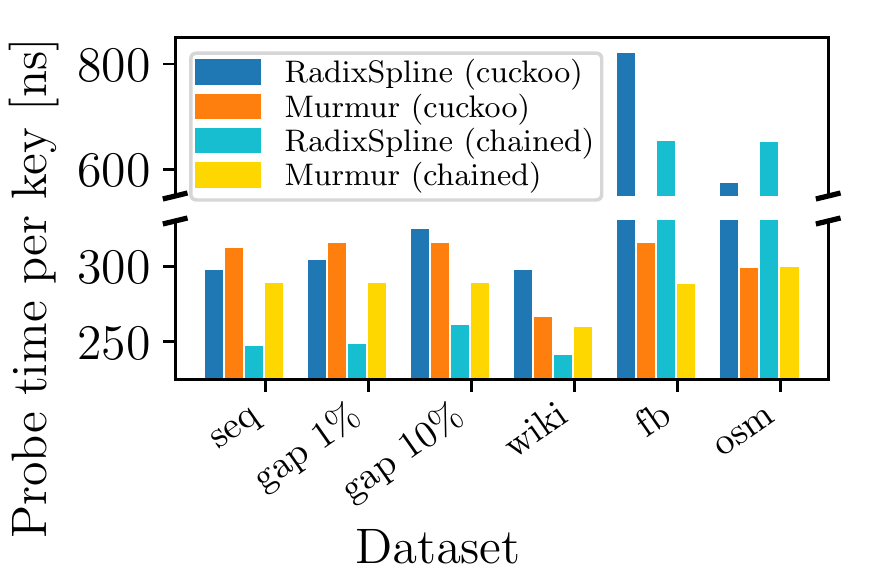}}
     \caption{Probe times comparison between Cuckoo and bucket chaining hashing.}
     \label{fig:probe_times_comparison}
\end{figure}

\noindent\textbf{Combined Probe Times.}  Figure~\ref{fig:probe_times_comparison} shows the probe times achieved by employing each of {\rs} and Murmur hashing inside both bucket chaining and Cuckoo hash tables on various datasets. We used bucket size of 4 for both tables. Also, biased kicking is used for constructing the Cuckoo table. As shown, for all datasets except \textit{fb} and \textit{osm}, bucket chaining with {\rs} is the best strategy. Cuckoo tables are generally slower than their chained counterparts.

%% file: appendix.tex
\input{appendix_models_for_less_collisions}
\input{appendix_models_as_fast_hash_functions}

%% file: appendix_models_for_less_collisions.tex
\section{Collision Analysis}
\label{app:collisions}
If we assume that $g_i$'s are generated from an independent and identically distributed (iid) variable with probability density function $f_{G}(x)$, then the expected number of empty slots $e$, after mapping keys to their hash outputs, is given by the formula below:
$$\mathbf{E}[e]=N  \cdot  \int_0^1 (1-x)\cdot f_{G}(x) dx $$
This equation was derived using the fact that if the gap value $x$ is less than one, then with probability $x$, a location boundary would fall between the two consecutive values.
This is because the location boundaries are separated by unit values and the probability of a random boundary falling in a gap of size $x$ is $x$.
The probability that no boundaries fall in a gap of size $x$ is $(1-x)$ and $(1-x)*f_{G}(x) dx$ represents the proportion collisions for gap values from [$x,x+dx$].  
This quantity is then integrated from $0$ to $1$, as consecutive keys with gaps beyond one don't collide.

%% file: appendix_models_as_fast_hash_functions.tex
\section{Algorithm for Hashing Via Learned Models}
\label{app:hash_via_learned_models_algo}

\begin{algorithm}[t]
\caption{{\sf Function} \textsc{HashViaLearnedModel} (Instances $s$, Keys $input$, KeysNum $N$, Output $hashes$)}
\begin{scriptsize}
\label{alg:rmi_hashing}
\begin{algorithmic}[1]
\State $done$ $\leftarrow$ 0 /* Flag to end hashing computation */
\State $state$ $\leftarrow$ \textsc{InitializeFSMInstances} ($s$) /* Initialize $s$ instances of a finite state machine*/
\While {$done < s$}
    \State $k$ = ($k$ == $s$) ? 0 : $k$
    \Switch {$state[k].stage$}
        \Case {P}: /* Predict using the root model, and prefetching next model parameters */
        \label{alg:state_1_start}
            \If {$i < N$}
                \State $state[k].vkey$ $\leftarrow$ \textsc{LoadKeyVec($input$, $i$)}
                \State $pred$ $\leftarrow$ \textsc{PredictNextLevelModelIndVec} ($state[k].vkey$, $root$) 
                \State $state[k].stage = H$
                \State $i += W$ /* $W$ is the vector width (Assume $N$ mod $W$) = 0*/
                \State \textsc{PrefetchModelParamsVec} ($pred$, $state[k].vparams$)
            \Else
                \State $state[k].stage = D$
                \State ++$done$
            \EndIf
        \EndCase
        \label{alg:state_1_end}
        \Case {H}: /* Perform actual hashing */
        \label{alg:state_2_start}
            \State $hashes$ $\leftarrow$ \textsc{HashKeysVec} ($state[k].vkeys$, $state[k].vparams$)
            \State $state[k].stage = P$ /* Initiate prefetching for a new set of keys in $P$*/
            \label{alg:state_2_end}
        \EndCase
    \EndSwitch
\EndWhile
\end{algorithmic}
\end{scriptsize}
\end{algorithm} 

Algorithm~\ref{alg:rmi_hashing} shows the pseudo code of our proposed batch-oriented hash function that combines {\amac} with vectorized learned models. The core idea is sample: for a vector of keys, we map the hashing computation generated by 2-levels learned model into an {\fsm} with two states, where the first state (Lines~\ref{alg:state_1_start} to~\ref{alg:state_1_end}) uses the root model to predict the index of the second level model, and prefetches its parameters, and the second state (Lines~\ref{alg:state_2_start} to~\ref{alg:state_2_end}) performs the actual hashing using the prefetched model parameters. The algorithm keeps interleaving multiple running instances of the {\fsm} till it finishes hashing all input keys. The logic in each state is completely implemented with {\simd} instructions.